\documentclass[12pt,showpacs,amsmath,amssymb,aps,pra]{revtex4}
\usepackage{bm}
\usepackage{amsfonts}
\usepackage{graphicx}
\begin{document}
\title{Quantum effects in the evolution of vortices in
the electromagnetic field}
\author{Tomasz Rado\.zycki}
\email{torado@fuw.edu.pl}
\affiliation{Department of Mathematical Methods in Physics, Faculty of
Physics, Warsaw University, Ho\.za 74, 00-682 Warsaw, Poland}

\begin{abstract}
We analyze the influence of electron-positron pairs creation on the 
motion of vortex lines in electromagnetic field. In our approach the 
electric and magnetic fields satisfy nonlinear equations derived from
the Euler-Heisenberg effective Lagrangian. We show that these
nonlinearities may change the evolution of vortices.
\end{abstract} 
\maketitle

\section{Introduction}

The phenomena of creation and evolution of vortices have always
attracted people's attention both in the past and in the present.  In
contemporary physics they gained particular interest since having been
experimentally observed in Bose-Einstein
condensates~\cite{and,mat,mad,che}. Vortices in superfluids, due to the
absence of viscosity, exhibit certain unconventional features like the
persistence of the whirl or its singular nature. The Bose-Einstein
condensate may be described by the nonlinear Schr\"odinger equation
(the Gross-Pitaevskii equation~\cite{gr,pit,fw}) satisfied by certain
macroscopic wave function. In that way one has been led to studying
vortices in quantum mechanics (QM). There is a striking resemblance
between dynamics of fluids and QM via the hydrodynamic formulation of
the latter~\cite{madel}. QM can, even in the linear version, serve as a
model theory for investigating the behaviour of vortices in
superfluids. Such studies, concerned with the dynamical as well as
topological aspects of vortex evolution in various configurations, both
in nonlinear~\cite{lund,rt1,rt2,kl1,kl2,sf,ar,gr1,gr2,izbb3} and in
linear~\cite{bbs,mb,md01,md02,bmrs} cases have recently been
undertaken.

Together with the attention paid to nonrelativistic QM the singular
solutions in other field theories as electromagnetism for instance have
been investigated~\cite{nye1,nye2,den,izbb1,ibb}. We will be concerned
with this question also in the present paper. While considering
vortices in fields corresponding to spinning particles one encounters
the problem, that the wave funcion has more than one component and the
condition $\psi({\bm r},t)=0$, leads to too many equations which cannot
be simultaneously satisfied.  One of the solutions of this problem for
electromagnetic field was proposed in Ref.~\onlinecite{izbb1}, where
vortex lines were defined by the null values of two relativistic
invariants: ${\cal S}$ and ${\cal P}$.  These two equations mean two
surfaces in the three-dimensional space.  Their intersection in general
may be a curve --- a vortex line. We will base our work on this
approach.

In the present work we would like to investigate how quantum effects
can influence the motion of the nodal lines of the electromagnetic wave
function~\cite{ibb2}. The term ``quantum'' is used here in the field
theoretical sense: Maxwell electrodynamics, as well as Schr\"odinger
wave mechanics, are classical from that point of view. We will
concentrate on the effect of electron-positron pairs creation in
electromagnetic fields. In order to construct the position and time
dependent wave function, we still need the classical equations for
electric and magnetic fields.  These equations are no longer linear
since pairs creation leads to the photon-photon interaction and Maxwell
equations are suplemented by additional terms which, in the lowest
approximation, are cubic in fields and quadratic in the fine-structure
constant (in general they might be also nonlocal). Although the
correction is small it can certainly influence the motion of the vortex
lines and particularly change their topology.

One should mention here that there exists also another type of quantum effects --- which remain beyond the concern of the present work --- connected not with the $e^+e^-$ content of the vacuum, but with fluctuations of the electromagnetic field itself. These effects, due to the nonzero vacuum expectation value of billinears in fields, lead to the smoothing of the vortex core (i.e. the line on which both invariants are equal to zero). In this case the core is no more singular. It is defined not by the condition $F^2=0$, which is not satisfied, but rather $F^2\approx 0$ , where $\approx$ means $|F|^2_{classical}<|F|^2_{vac.fluct.}$~\cite{mvbmrd}, where $F$ is the Riemann-Silberstein vector spoken more of in Sections II and III.

As a starting point we choose the Euler-Heisenberg (E-H)
Lagrangian~\cite{eh,js} describing, in the lowest order, the dynamics
of classical electromagnetic fields with vacuum polarization effects
taken into account. Field equations obtained from this effective theory
in Section II, exhibit solutions containing vortex lines, the evolution
of which may be viewed and compared to that obtained from the classical
Maxwell equations. In this work we analyze two such cases. Both are
chosen from~\cite{izbb1} to make the comparison of the results in our
works very easy. Our results are presented in Section III.

The main practical problem in this investigation comes from the fact
that quantum corrections are, in general, small and it is very hard to
see them on a drawing. The choice of examples considered in our work
from among those of Ref.~\onlinecite{izbb1} is dictated just by the
criterion of quantum effects being visible. It is clear that they are
noticeable not by analyzing or measuring the precise shape of a vortex
line --- the slight deviation of which from that obtained in classical
theory surely does occur --- but rather by observing ``to be or not to
be'' effects or topological ones.

There are two limitations which cause some of the results of this work
to be qualitative rather than quantitative. Firstly they are obtained
within perturbative regime. This regime means that the electromagnetic
field may not be too strong and its strength is limited by the
condition of $\alpha|{\bm F}|^2/m^4$ being small. Fields considered
here are polynomial, so this requirement means that the evolution
should not go beyond certain limited space-time region. However,
close to the vortex line, similarly to the situation in
Gross-Pitaevskii equation, the terms quadratic in ${\cal
S}$ and ${\cal P}$ in the Hamiltonian~(\ref{ham}) become small, even
for large values of electromagnetic fields, and perturbative
calculation is again well justified.

Secondly we have to remember that that E-H Lagrangian describes only
slowly varying fields, for which the nonlocality may be neglected.
Their relative change at a distance of the Compton wavelength of the
electron should be small. In view of that the E-H effective Lagrangian
is treated in our work as a certain nonlinear model of the true theory
of electromagnetic fields obtained form QED without real charges. An
another interesting model in this context constitutes the Born-Infeld
electrodynamics~\cite{bi}. One should, however, have in mind that even small
corrections, coming from weak fields can change the topology of
vortices.

\section{Field equations}

The Euler-Heisenberg Lagrangian~\cite{eh,js}, which accounts
for the vacuum polarization processes in the lowest approximation, has
the following form
\begin{equation}\label{lagr}
{\cal L}({\bm r},t)={\cal S}({\bf
r},t)+\frac{2\alpha^2}{45 m^4}\left[4{\cal S}({\bm r},t)^2+7{\cal P}({\bm 
r},t)^2\right]\; ,
\end{equation}
where $\cal S$ and $\cal P$ denote the two Poincar\'e invariants formed
of electromagnetic fields
\begin{equation}\label{invar}
{\cal S}=\frac{1}{2}\left({\bm E}^2-{\bm B}^2\right)\; ,\hspace{2cm}
{\cal P}={\bm E}\cdot{\bm B}
\end{equation}
and $\alpha$ and $m$ are fine-structure constant and electron mass
recpectively. 
The canonical variables are electric and magnetic inductions: ${\bm D}$
and ${\bm B}$, where the former plays the role of canonical momentum and
the latter of position~\cite{bi, izbb2}. In this picture the electric
field strength 
${\bm E}$ in the Lagrangian~(\ref{lagr}) corresponds to velocity. 
The field equations, we will need for our purpose, are
the canonical Hamilton equations
\begin{subequations}\label{caneqs}
\begin{eqnarray}
\dot{\bm D}({\bm r},t)&=&{\bm \nabla}\times\frac{\partial {\cal H}({\bm 
r},t)}{\partial {\bm B}({\bm r},t)}\; ,\\
\dot{\bm B}({\bm r},t)&=&-{\bm \nabla}\times\frac{\partial {\cal H}({\bm 
r},t)}{\partial {\bm D}({\bm r},t)}\; ,
\end{eqnarray}
\end{subequations}
where ${\cal H} ({\bm r},t)$ denotes the Hamiltonian density. 
To find the explicit form of~(\ref{caneqs}) we have to perform the Legendre
transform and pass from $\cal L$ to $\cal H$.
The canonical momentum is, as always, defined as a derivative of the
Lagrangian over velocity 
\begin{equation}\label{ddef}
{\bm D}({\bm r},t)=\frac{\partial {\cal L} ({\bm r},t)}{\partial {\bm E}({\bm 
r},t)}=\frac{\partial {\cal L} ({\bm r},t)}{\partial {\cal S}({\bm 
r},t)}{\bm E}({\bm r},t)+\frac{\partial {\cal L} ({\bm r},t)}{\partial
{\cal P}({\bm r},t)}{\bm B}({\bm r},t)\; ,
\end{equation}
which gives
\begin{equation}\label{d}
{\bm D}({\bm r},t)=\left[1+\frac{16\alpha^2}{45m^4}{\cal
S}({\bm r},t)\right]{\bm E}({\bm r},t)+
\frac{28\alpha^2}{45m^4}{\cal
P}({\bm r},t){\bm B}({\bm r},t)\; .
\end{equation}

This kind of equation usually bears the name of a constitutive equation.
It reflects the nontrivial structure of the medium. In the present case
this medium is the quantum field theory vacuum with its polarizability
via electron-positron pairs creation and anihilation.  We now need to
invert this equation and express velocity ${\bm E}$ in terms of
canonical variables ${\bm D}$ and ${\bm B}$. Since our initial
Lagrangian~(\ref{lagr}) is given only in one loop approximation
($\alpha^2$) then our further calculations may be led up to this order
too. We can therefore postulate ${\bm E}({\bm r},t)$ in the form
\begin{equation}\label{edef}
{\bm E}({\bm r},t)=\left[1+\alpha^2{\cal K}({\bm 
r},t)\right]{\bm D}({\bm r},t)+ 
\alpha^2 {\cal M}({\bm r},t){\bm B}({\bm r},t)\; ,
\end{equation}
where quantities ${\cal K}({\bm r},t)$ and ${\cal M}({\bm r},t)$ are to be
determined. Substituting~(\ref{edef}) into~(\ref{d}), neglecting terms
of the order higher than $\alpha^2$, and comparing coefficients
multiplying vectors  ${\bm D}$ and ${\bm B}$ we find
\begin{subequations}
\begin{eqnarray}\label{km}
{\cal K}({\bm r},t)&=&-\frac{16}{45 m^4}\left[{\bm D}({\bm r},t)^2-{\bf
B}({\bm r},t)^2\right]\; ,\\
{\cal M}({\bm r},t)&=&-\frac{28}{45 m^4}{\bm D}({\bm r},t)\cdot{\bf
B}({\bm r},t)\; . 
\end{eqnarray}
\end{subequations}

The Hamiltonian density may be now found as
\begin{equation}\label{hamdef}
{\cal H}({\bm r},t)={\bm E}({\bm r},t)\cdot {\bm D}({\bm r},t)-{\cal L}({\bm
r},t)\; ,
\end{equation}
where ${\bm E}$ in the whole above expression should be eliminated in favor
of ${\bm D}$ and ${\bm B}$, according to the relations~(\ref{edef},\ref{km}). The explicit
form of ${\cal H}$ is then
\begin{equation}\label{ham}
{\cal H}({\bm r},t)=\frac{1}{2}\left[{\bm D}({\bm r},t)^2+{\bm B}({\bm
r},t)^2\right]-\frac{2\alpha^2}{45 m^4}\left[{\bm D}({\bm r},t)^2-{\bm B}({\bm
r},t)^2\right]^2-\frac{14\alpha^2}{45 m^4}\left[{\bm D}({\bm
r},t)\cdot{\bm B}({\bm r},t)\right]^2\; .
\end{equation}
Now we are in a position to write down the equations~(\ref{caneqs})
in an explicit form 
\begin{subequations}\label{dbeqs}
\begin{eqnarray}
\dot{\bm D}({\bm r},t)={\bm \nabla}\times\bigg\{{\bm B}({\bm r},t)&&\left[
1+\frac{8\alpha^2}{45 m^4}\left({\bm D}({\bm r},t)^2-{\bm B}({\bm
r},t)^2\right)\right]\\*
&&-\frac{28\alpha^2}{45 m^4}{\bm D}({\bm r},t)\left[{\bm D}({\bm
r},t)\cdot{\bm B}({\bm r},t)\right]\bigg\}\; ,\nonumber\\
\dot{\bm B}({\bm r},t)=-{\bm \nabla}\times\bigg\{{\bm D}({\bm r},t)&&\left[
1-\frac{8\alpha^2}{45 m^4}\left({\bm D}({\bm r},t)^2-{\bm B}({\bm
r},t)^2\right)\right]\\*
&&-\frac{28\alpha^2}{45 m^4}{\bm B}({\bm r},t)\left[{\bm
D}({\bm r},t)\cdot{\bm B}({\bm r},t)\right]\bigg\}\; ,\nonumber
\end{eqnarray}
\end{subequations}

Introducing two complex vectors ${\bm F}_{\pm}({\bm r},t)$ according
to the relation
\begin{equation}\label{fdef}
{\bm F}_{\pm}({\bm r},t)=\frac{1}{\sqrt{2}}\left({\bm D}({\bm r},t)\pm
i{\bm B}({\bm r},t)\right)
\end{equation}
we can rewrite the equations~(\ref{dbeqs}) in the form
\begin{subequations}\label{fpm}
\begin{eqnarray}\label{fp}
\dot{\bm F}_+({\bm r},t)&=&-i{\bm \nabla}\times{\bm F}_+({\bm r},t)+\frac{2 i
\alpha^2}{45 m^4}{\bm \nabla}\times\left[{\bm F}_-({\bm r},t)\left(11{\bm
F}_+({\bm r},t)^2-3{\bm F}_-({\bm r},t)^2\right)\right]\; ,\label{fplus}\\
\dot{\bm F}_-({\bm r},t)&=&i{\bm \nabla}\times{\bm F}_-({\bm r},t)+\frac{2 i
\alpha^2}{45 m^4}{\bm \nabla}\times\left[{\bm F}_+({\bm r},t)\left(11{\bm
F}_-({\bm r},t)^2-3{\bm F}_+({\bm r},t)^2\right)\right]\label{fminus}\; .\label{fm}
\end{eqnarray}
\end{subequations}

In the classical case the right hand sides of~(\ref{fpm}) reduce to the
first terms only and the two equations for ${\bm F}_{\pm}$ decouple
from each other. This is not the case in the presence of a nonlinear
medium.

The evolution takes place in an empty space, without real charges, so
${\bm F}_{\pm}({\bm r},t)$ have to satisfy the conditions
\begin{equation}\label{empty}
{\bm \nabla}\cdot{\bm F}_{\pm}({\bm r},t)=0\; .
\end{equation}
By applying gradient to both sides of~(\ref{fpm}) it can easily be seen
that ${\bm \nabla}\cdot{\bm F}_{\pm}({\bm r},t)$ are constant in time
and it is sufficient to impose the conditions~(\ref{empty}) at time
$t=0$.

\section{Evolution of exemplary vortices}

In the present section we would like to show how quantum effects
connected with pairs creation, influence the evolution of vortices in
the electromagnetic field. From
among the configurations of vortex lines considered
in~\cite{izbb1} we have chosen two, for which the comparison can
most easily be done and the effects are clearly visible. They are the
situations presented in Figures 1 and 2 of~\cite{izbb1}: the
motion of the vortex ring and the creation and further evolution of
initially linear vortex-antivortex configuration, i.e. two vortices of
opposite whirl. 

Vortex lines in quantum mechanics are usually defined by the behaviour
of the wave function of the system.  In hydrodynamics vortices appear
in the regions of space where ${\bm \nabla}\times{\bm v}({\bm r},t)\neq
0$, where ${\bm v}({\bm r},t)$ is the local fluid velocity. In QM, in
its hydrodynamic formulation~\cite{mad}, the role of the fluid is
played by the distrubution of probability. The velocity field, being
proportional to the gradient of the phase of the wave function, can
have nonvanishing curl only where this phase is singular. This in turn
means the vanishing of the wave function, i.e. the simultaneous
vanishing of its real and imaginary parts. In that way we are led to
the conclusion that, in general, vortices have the character of the
curves (evolving in time) costituting the intersection of two surfaces
defined by the requirement $\psi({\bm r},t)=0$.  

As it was proposed in~\cite{izbb1} one can introduce in electrodynamics,
in place of $\psi$, a similar object, the vanishing of which may serve as the
definition for the vortex lines. This object is ${\bm F}({\bm r},t)^2$,
where ${\bm F}({\bm r},t)=\frac{1}{\sqrt{2}}\left({\bm D}({\bm r},t)+
i{\bm B}({\bm r},t)\right)$. As argued~\cite{ibb2}
the quantity ${\bm F}$ is worthy of being called a ``photon wave function''.
In the case considered in the present work, photons move in the
polarizable medium, and what is more, the nonlinear one. Already in the
linear (but inhomogeneous) medium one is forced to define
the wave function as an extention of ${\bm
F}$ through the introduction of upper and lower components~\cite{ibb2}
defined by in~(\ref{fdef})
\begin{equation}\label{wavef}
{\cal F}({\bm r},t)=\left(\begin{array}{l} {\bm F}_+({\bm r},t)\\
{\bm F}_-({\bm r},t) 
\end{array}\right)\; , 
\end{equation}
This allows one to give the set of coupled equations the form of one,
linear, Schr\"odinger-type equation for the wave function ${\cal
F}({\bm r},t)$. In the quantum case the linearity is inevitably lost,
but the definition of vortex lines, by the requirements ${\cal S}({\bm
r},t)=0$ and ${\cal P}({\bm r},t)=0$, seems to be universal
(following~\cite{izbb1} this kind of singular lines has recently been
called ``Riemann-Silberstein'' vortices~\cite{mbe,kai}).  Therefore, in
the full analogy with~\cite{izbb1}, we choose as a basic object the
quantity
\begin{equation}\label{f2}
{\bm F}_+^2=\frac{1}{2}({\bm D}^2-{\bm B}^2)+i{\bm D}\cdot{\bm B}\; .
\end{equation}
The condition ${\bm F}_+({\bm r},t)^2=0$ is naturally equivalent to the
choice ${\bm F}_-({\bm r},t)^2=0$.

\subsection{Vortex ring}

The first configuration considered in~\cite{izbb1} is defined by 
\begin{equation}\label{fbb1}
{\bm f}^{(a)}({\bm r},t)=\left(y+it,z-a+i(a+t),x+it\right)\; .
\end{equation}

This ``wave function'' satisfies the Maxwell equations and describes
the evolution of a single vortex in the form of a swinging ring with
varying radius. In order to see in an easy way how quantum (nonlinear)
terms in~(\ref{fpm})
influence this evolution, we will choose the solution ${\bm F}_+({\bm
r},t)$ of~(\ref{fpm}) which is identical to~(\ref{fbb1}) at $t=0$. This
solution (up to $\alpha^2$) has the form
\begin{equation}\label{sola}
{\bm F}_+^{(a)}({\bm r},t)={\bm f}^{(a)}({\bm r},t) 
+t^3\cdot{\bm \alpha}({\bm r})+t^2\cdot{\bm \beta
}({\bm r})+t\cdot{\bm \gamma}({\bm r})\; ,
\end{equation}
where vector funtions ${\bm\alpha}({\bm r})$,
${\bm\beta}({\bm r})$ and ${\bm\gamma}({\bm r})$ are given by
\begin{subequations}\label{abca}
\begin{eqnarray}
{\bm \alpha}({\bm r})&=&-\frac{128 i \alpha^2}{135 m^4}(1,1,1)\; ,\\
{\bm\beta}({\bm r})&=&\frac{8 \alpha^2}{3
m^4}\left(\frac{1}{3}(z-a)-\frac{2}{5}y-\frac{i}{3}a,
\frac{2}{5}(a-z)+\frac{1}{3}x-\frac{i}{3}a,-\frac{2}{5}x+\frac{1}{3}y
\right)\; ,\\ 
{\bm\gamma}({\bm r})&=&\frac{8 \alpha^2}{3
m^4}\left(\frac{2}{3}a(z-a)-\frac{i}{15}(11a^2-12az+6z^2),
-\frac{2i}{5}x^2,-\frac{2i}{5}y^2
\right)\; .
\end{eqnarray}
\end{subequations}

\begin{figure}[ht]{
\includegraphics[width=14cm]{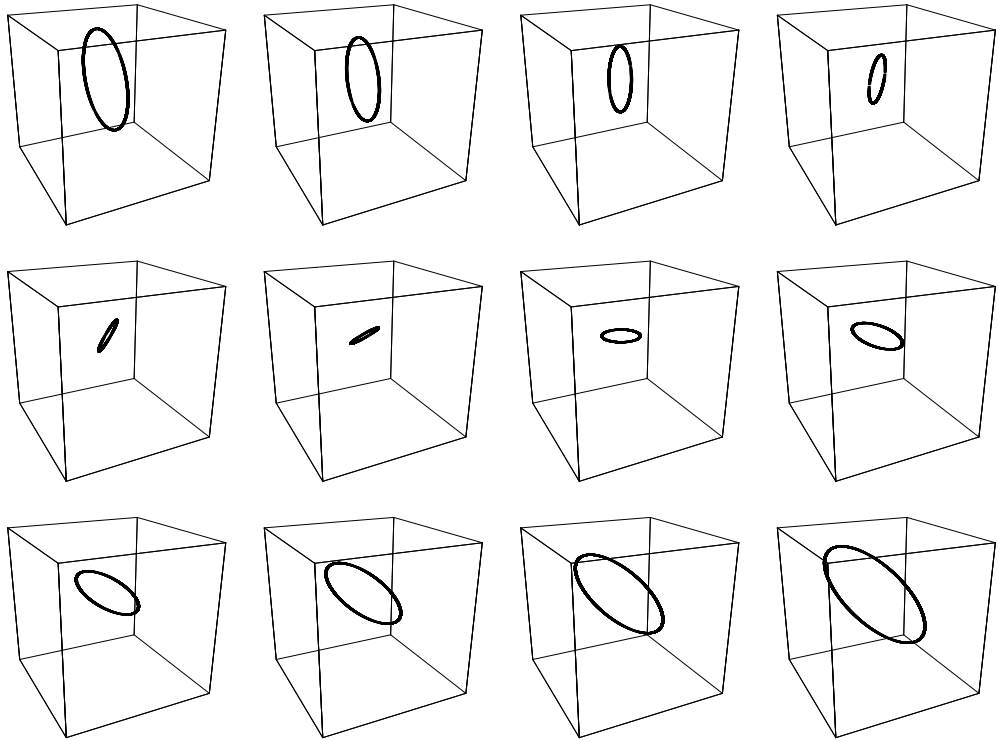}
\caption{The evolution of the vortex ring starting from $t=-1.8$ to
$t=1.5$. The scale on the axes is such that the frame covers the region
$-4<x,y,z<4$.} \label{fig:ring}}
\end{figure}

In Figure~\ref{fig:ring} we show the evolution of the vortex line constituting
the intersection of two surfaces $\Re {\bm F}_+^{(a)}({\bm r},t)^2=0$ and
$\Im {\bm F}_+^{(a)}({\bm r},t)^2=0$ with ${\bm F}_+^{(a)}({\bm
r},t)$ defined by~(\ref{sola}) and~(\ref{abca}). 

\begin{figure}[ht]{
\includegraphics[width=14cm]{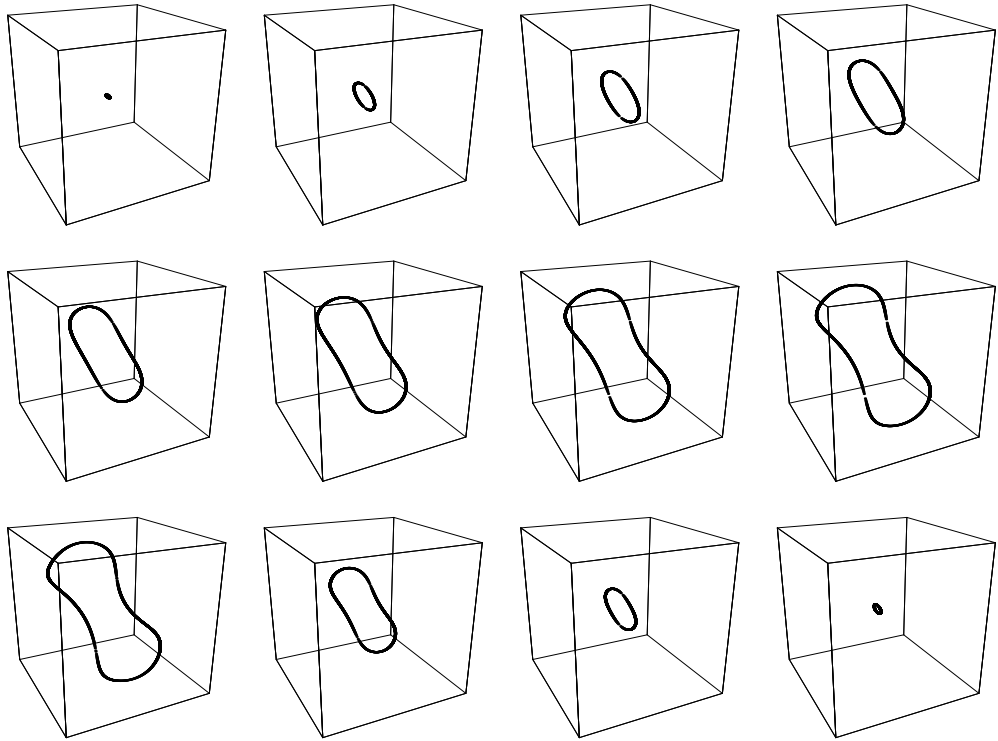}
\caption{The evolution of the vortex ring for larger times: from $t=1.5$
to $t=99.3$. The first frame is identical to the last one of
Figure~\ref{fig:ring}, but now the scale of the axes is modified to
$-100<x,y,z<100$.} \label{fig:ringq}}
\end{figure}

For simplicity both parameters $a$ and $m$ are set equal to unity on
this, as well as on the following plots. The evolution extends in time
from $t=-1.8$ to $t=1.5$ and exhibits identical character to that
of~\cite{izbb1}: the swinging vortex ring, preserving its circular
character, decreases to certain minimal value of radius, and then
starts to increase. Quantum effects do not manifest themselves in this
domain of space and time. In the classical case, hovever, the expansion
of a ring will last forever, which can easily be seen from the two
equations given in~\cite{izbb1}
\begin{subequations}
\begin{eqnarray}\label{vorta}
x^2+y^2+(z-a)^2-a^2-2at-3t^2&=&0\; ,\\
2az+2t(x+y+z-a)-2a^2&=&0\; .
\end{eqnarray}
\end{subequations}
The former represents the sphere of a fixed center in the point
$(0,0,a)$ and of constantly increasing radius (for positive $t$). The
latter, rewritten in the form $x+y+(1+a/t)z=a(1+a/t)$, tends to the
motionless plane $x+y+z=a$ passing through the center of the sphere.
Their intersection will surely be the expanding ring. As we see in
Figure~\ref{fig:ringq}, this ceases to be true in the quantum case. 

Due to the nonlinearity introduced by quantum effects two new phenomena
appear. Firstly the vortex ring starts to deviate, for intermediate times, 
from its regular, circular character. Secondly it is no longer
constantly expanding. On the contrary, after reaching certain
maximal extention it starts to decrease down to its complete
disappearance, if we drew also frames for larger times.

If we traced the vortex evolution even further in time (certainly
far beyond the applicability of the perturbative methods) we would
observe the complicated system of vortices approaching from
``infinity''.

\subsection{Vortex-antivortex}

The second situation corresponds to the case (b) of~\cite{izbb1}
\begin{equation}\label{fbb2}
{\bm f}^{(b)}({\bm r},t)=\left(y+t,a-i(z+a-t),x+it\right)\; .
\end{equation}

\begin{figure}[ht]{
\includegraphics[width=14cm]{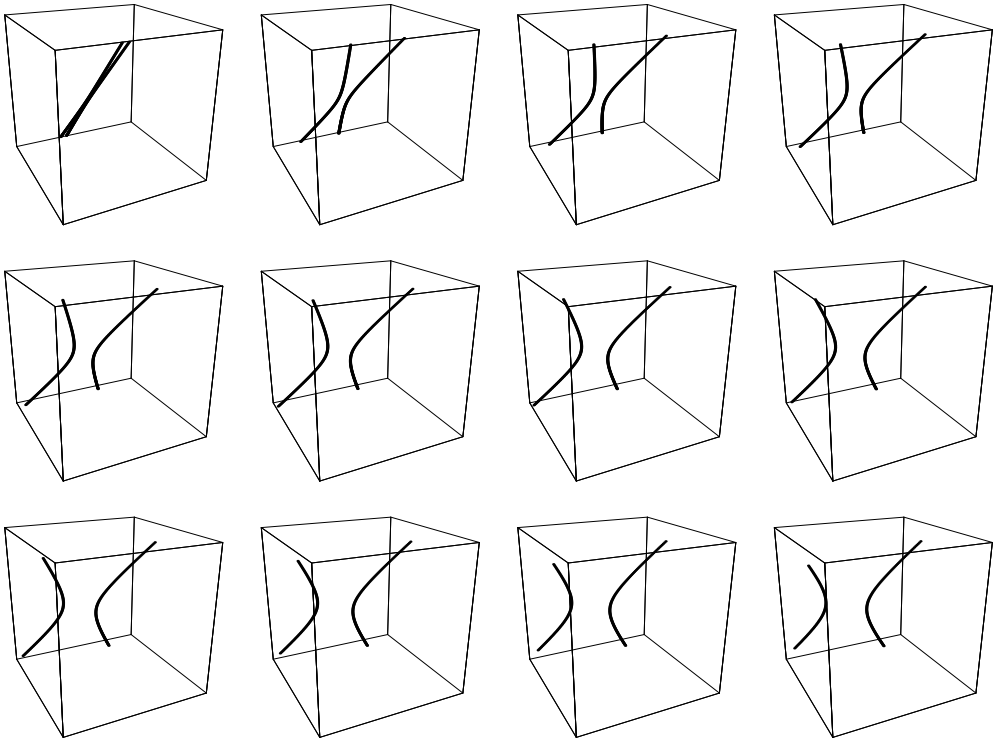}
\caption{The evolution of the system of two ``antiparallel'' vortices for
time between $t=1$ and $t=1.55$. The units on the axes are such that
each frame represents the cube $-4<x,y,z<4$.} \label{fig:vav}}
\end{figure}

The above function was shown to describe the configuration of two
vortices which initially are antiparallel straight lines (we call them vortex
and antivortex). They are born at $t=a$ and then they separate and
deform.  The solution of~(\ref{fp}) which is identical to ${\bm
f}^{(b)}$ at $t=0$ has (again up to $\alpha^2$) the form 
similar to~(\ref{sola})
\begin{equation}\label{solb}
{\bm F}_+^{(b)}({\bm r},t)={\bm f}^{(b)}({\bm r},t) 
+t^3\cdot{\bm \alpha}({\bm r})+t^2\cdot{\bm \beta
}({\bm r})+t\cdot{\bm \gamma}({\bm r})\; ,
\end{equation}
but now with
\begin{subequations}\label{abcb}
\begin{eqnarray}
{\bm \alpha}({\bm r})&=&-\frac{8 \alpha^2}{15
m^4}\left(2,\frac{17i}{9},\frac{17i}{9}\right)\; ,\\ 
{\bm\beta}({\bm r})&=&\frac{8 \alpha^2}{15
m^4}\left(2(z-y+a)+\frac{5i}{3}a,
\frac{5}{3}(x-a)+2i(z+a),-2x+2iy
\right)\; ,\\ 
{\bm\gamma}({\bm r})&=&-\frac{8 \alpha^2}{15
m^4}\left(\frac{11}{3}a^2+4az+2z^2+\frac{10i}{3}a(z+a),
2ix^2,2iy^2 \right)\; .
\end{eqnarray}
\end{subequations}
The evolution of this vortex configuration is presented in
Figure~\ref{fig:vav}. Again we have put $a=1$ and $m=1$.

We see in general the same motion as that found in~\cite{izbb1} except
one difference visible in the first frame. In~\cite{izbb1} the two
straight, antiparallel vortices spring up at $t=a$ (this time
corresponds to the first frame) as exactly overlapping. No vortices
exist for $t$ between $-a$ and $a$. In the case of Figure~\ref{fig:vav}
the vortices in the first frame a slightly shifted and of different
slope.  This is a result of the influence of the nonlinear (quantum)
terms in~(\ref{fpm})\footnote{This difference is truely of quantum
origin and is not connected with the mistaken sign in the formula (19b)
in~\cite{izbb1}, which is only a literal error (private
communication).}. We recall that the vectors $F^{(b)}({\bm r},t)$ and
$f^{(b)}({\bm r},t)$ are synchronized for $t=0$ and not for $t=a$.
This small shift and deformation are then consequences of the quantum
correction to the evolution for $0<t<a$.

\begin{figure}[ht]{
\includegraphics[width=14cm]{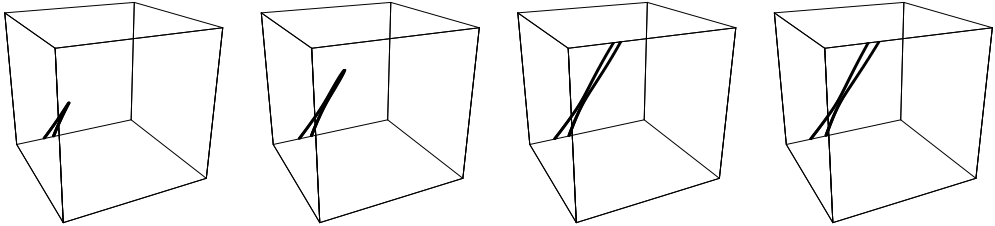}
\caption{The appearance of the system of vortices of Figure~\ref{fig:vav}. The
frames correspond to times just before $t=a$. Now the scale on the $y$
axis is changed to make the splitting of vortices easily visible:
$-4<x,z<4$ and $-1.5<y<0.5$. } \label{fig:vavq}}
\end{figure}

In~\cite{izbb1} the system of vortices is born at $t=a$, but in our
case they do not overlap and consequently must have appeared earlier.
It is therefore interesting to take a step back in time and see how
these vortices arise in the quantum case. This is shown in
Figure~\ref{fig:vavq}.

\begin{figure}[ht]{
\includegraphics[width=14cm]{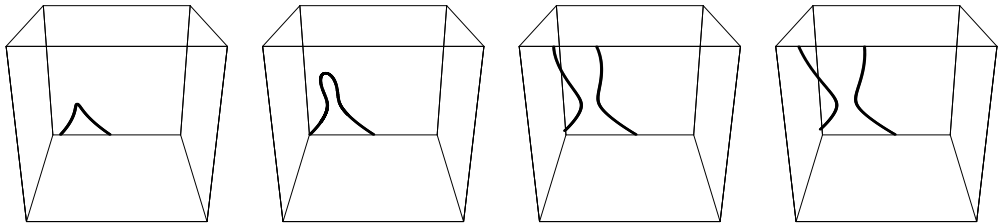}
\caption{The appearance of the system of vortices of
Figure~\ref{fig:vav} seen now from the viewpoint other than that of
Figure~\ref{fig:vavq} and with $y$-axis rescaled even more.} 
\label{fig:vavqext}}
\end{figure}

Figure~\ref{fig:vavq} brings to light the essential change: the two
independent vortices in the classical case, or rather vortex and
antivortex, become the two fractions of the same, tightly bent, vortex
line, when quantum corrections are taken into account. Their sudden
creation turns out now to be a motion during which this single vortex
line simply enters into the observation region and is being deformed.
One might expect this kind of effects --- that could be called
topological effects --- together with the smoothing of the evolution,
to be the most typical ones introduced by the nonlinearity of the
quantum equations. To make the effect more visible we present it again
on Figure~\ref{fig:vavqext}, now seen from another viewpoint.

We would like also to emphasize that the above phenomena take place for
electromagnetic fields weak enough to remain in full agreement with the use
of perturbation theory.

Yet another difference, however not visible on Figures~\ref{fig:vavq}
and~\ref{fig:vavqext},  is the
slight deviation of the system of vortex lines from their planar
character. In the clasical case the vortex lines arose as the
intersection of a plane with a certain surface, and therefore all vortices
have to lie forever in one plane. This is no longer true in the quantum
case.

\section{Summary}

In the present paper we analyzed the influence of the nonlinear,
quantum terms in the Maxwell equations on the evolution of vortex
lines. By making the comparison with the results obtained earlier in
the classical case~\cite{izbb1} we found that this evolution may be
changed in a visible and essential way. In the first considered
configuration of the constantly expanding vortex ring, our calculations
show that quantum corrections may lead to deformation and to
disappearance of this ring. In the second case of two linear and
antiparallel vortices of the infinite size, which are suddenly created,
we show how the process of this ``creation'' looks like, and that the
two independent vortices (in the classical case) turn out to be just
different fractions of the same vortex curve. This kind of topological
changes might be expected as a result of nonlinearity introduced by
vacuum polarization.

The present analysis has certain limitations which come both from its
perturbative character and from the ``low frequency'' approximation
which allowed one to derive the E-H Lagrangian. It can, however, serve
as a qualitative picture of what type of phenomena may be introduced by
the quantum effects. One is still very far from constructing the
nonperturbative solutions of Quantum Electrodynamics, which would be
deprived of the above limitations, and therefore it might be also
interesting to consider the evolution of nodial lines in certain exact
nonlinear theory as Born-Infeld electrodynamics. However, in this case,
one cannot expect to find the polynomial solutions as
given by~(\ref{sola},\ref{solb}) and only numerical calculations
come into play. This situation is similar to that in the
nonlinear quantum mechanics.

At the end we would like to note that although the observed deformation and evolution of vortices have their roots in the quantum nature of the vacuum, similar structures may also appear in classical and linear fields by the appropriate perturbation of the vortex configurations. Both the deviation of a vortex ring  from the planar character as well as the occurrence of a `hairpin'-shaped vortex, similar to that of Figure~\ref{fig:vavqext}, are known in optical difraction~\cite{bnw}.


\begin{thebibliography}{}
\bibitem{mat} M.R. Matthews \textit{et al.}, Phys. Rev. Lett. {\bf 83},
2498(1999). 
\bibitem{mad} K.W. Madison \textit{et al.}, Phys. Rev. Lett. {\bf 84},
806(2000). 
\bibitem{che} F. Chevy, K.W. Madison and J. Dalibard, Phys. Rev. Lett.
{\bf 85}, 2223(2000).
\bibitem{and} B.P. Anderson \textit{et al.}, Phys. Rev. Lett. {\bf 85},
2857(2000).
\bibitem{gr} E.P. Gross, N. Cim. {\bf 20}, 454(1961).
\bibitem{pit} L.P. Pitaevskii, Zh. Eksp. Teor. Fiz. {\bf 40}, 646(1961)
[Sov. Phys. JETP {\bf 13}, 451(1961)].
\bibitem{fw} A.L. Fetter and J.D. Walecka, \textit{Quantum Theory
of Many-Particle Systems}, McGraw-Hill, New York 1971.
\bibitem{madel} O. Madelung, Z. Phys. {\bf 40}, 342(1926).
\bibitem{lund} F. Lund, Phys. Lett. A {\bf 159}, 245(1991).
\bibitem{rt1} S. Rica and E. Tirapegui, Physica D {\bf 61}, 246(1992).
\bibitem{rt2} S. Rica and E. Tirapegui, Phys. Rev. Lett. {\bf 64},
878(1990). 
\bibitem{kl1} J. Koplik and H. Levine, Phys. Rev. Lett. {\bf 71},
1375(1993). 
\bibitem{kl2} J. Koplik and H. Levine, Phys. Rev. Lett. {\bf 76},
4745(1996).
\bibitem{sf} A.A. Svidzinsky and A.L. Fetter, Phys. Rev. A {\bf 62},
063617(2000). 
\bibitem{ar} A. Aftalion and T. Riviere, Phys. Rev. A {\bf 64}, 043611(2001).
\bibitem{gr1} J.J. Garcia-Ripoll and V.M. P\'erez-Garcia, Phys. Rev. A
{\bf 64}, 053611(2001).
\bibitem{gr2} J.J. Garcia-Ripoll \textit{et al.}, Phys. Rev. Lett.
{\bf 87}, 140403(2001).
\bibitem{izbb3} I. Bia{\l}ynicki-Birula and Z. Bia{\l}ynicka-Birula,
Phys. Rev. A {\bf 65}, 014101(2001).
\bibitem{bbs} I. Bia{\l}ynicki-Birula, Z. Bia{\l}ynicka-Birula and C.
\'Sliwa, Phys. Rev. A {\bf 61}, 032110(2000).
\bibitem{mb} M. Berry, Found. Phys. {\bf 31}, 659(2001).
\bibitem{md01} M.V. Berry and M.R. Dennis, Proc. Roy. Soc. Lond. A
{\bf 457}, 2251(2001).
\bibitem{md02} M.V. Berry and M.R. Dennis, J. Phys. A
{\bf 34}, 8877(2001).
\bibitem{bmrs} I. Bia{\l}ynicki-Birula \textit{et al.}, Acta. Phys.
Pol. A Suppl. {\bf 100}, 29(2001). 
\bibitem{nye1} J.F. Nye, Proc. Roy. Soc. Lond. A {\bf 387}, 105(1983).
\bibitem{nye2} J.F. Nye, Proc. Roy. Soc. Lond. A {\bf 389}, 279(1983).
\bibitem{den} M.R. Dennis, \textit{Topological Singularities in Wave
Fields}, Ph.D. Thesis, Bristol 2001.
\bibitem{izbb1} I. Bia{\l}ynicki-Birula and Z. Bia{\l}ynicka-Birula,
Phys. Rev. A {\bf 67}, 062114(2003).
\bibitem{ibb} I. Bia{\l}ynicki-Birula, to appear in J. Opt. A.
\bibitem{ibb2} Iwo Bia{\l}ynicki-Birula, Prog. Opt. {\bf 36}, 245(1996)
and references therein.
\bibitem{mvbmrd}  M.V. Berry and M.R. Dennis, to appear in J. Opt. A.
\bibitem{eh} H. Euler and W. Heisenberg, Z. Phys. {\bf 98}, 714(1932); 
\bibitem{js} J. Schwinger, Phys. Rev. {\bf 82}, 664(1951);
\bibitem{bi} M. Born and L. Infeld, Proc. Roy. Soc. {\bf A 147},
522(1934); {\bf A 150}, 141(1935).
\bibitem{izbb2} I. Bia{\l}ynicki-Birula and Z. Bia{\l}ynicka-Birula,
\textit{Quantum Electrodynamics}, Pergamon, Oxford 1975.
\bibitem{mbe} M.V. Berry, to appear in J. Opt. A.
\bibitem{kai} G.J. Kaiser, to appear in J. Opt. A.
\bibitem{bnw} M.V. Berry, J.F. Nye, F.J. Wright, Phil. Trans. Roy. Soc. {\bf A 291}, 453(1979).
\end{thebibliography}
\end{document}